\documentclass[final,3p,two column,times]{elsarticle}
\usepackage{setspace}
\usepackage[latin9]{inputenc}
\usepackage{color}
\usepackage{array}
\usepackage{float}
\usepackage{textcomp}
\usepackage{multirow}
\usepackage{amsmath}
\usepackage{amsthm}
\usepackage{amssymb}
\usepackage{setspace}
\usepackage{graphics}
\usepackage{graphicx}

\makeatletter

\DeclareRobustCommand{\greektext}{%
  \fontencoding{LGR}\selectfont\def\encodingdefault{LGR}}
\DeclareRobustCommand{\textgreek}[1]{\leavevmode{\greektext #1}}
\DeclareFontEncoding{LGR}{}{}
\DeclareTextSymbol{\~}{LGR}{126}
\providecommand{\tabularnewline}{\\}
\@ifundefined{date}{}{\date{}}

\usepackage{graphics}\usepackage{amsthm}\usepackage{lineno}

\journal{Nuclear Instruments and Methods in Physics Research A}

\newcommand{\labr}{LaBr$_3$(Ce)}

\newcommand{\baf}{BaF$_2$}
\def\nuc#1#2{\relax\ifmmode{}^{#1}{\protect\text{#2}}\else${}^{#1}$#2\fi}

\begin{document}
\begin{frontmatter}


\title{Enhanced time response of 1-inch LaBr$_3$(Ce) crystals by 
leading edge and constant fraction techniques}

\author[UCM]{V. Vedia\corref{cor1}}
\ead{mv.vedia@ucm.es}
\ead[url]{http://nuclear.fis.ucm.es}
\author[UCM,NCBJ]{H. Mach\fnref{deceased}}
\author[UCM]{L.M. Fraile}
\author[UCM]{J.M. Ud\'{i}as}
\author[Sof]{S. Lalkovski\fnref{Surrey}}
\cortext[cor1]{Corresponding author}
\fntext[deceased]{During the preparation of the final version of this manuscript, our
colleague Henryk Mach suddenly passed away. Henryk was an inspirational character for us and for
several generations of scientists. He will be greatly missed.}
\fntext[Surrey]{Present addresses: Department of Physics, Faculty of Engineering and Physical Sciences, University of Surrey
Guildford, GU2 7XH, United Kingdom}

\address[UCM]{Grupo de F\'isica Nuclear, Facultad de CC. F\'isicas, Universidad Complutense, CEI Moncloa, ES-28040 Madrid, Spain}
\address[NCBJ]{National Centre for Nuclear Research, Division for Nuclear Physics, BP1, PL-00-681 Warsaw, Poland}
\address[Sof]{Faculty of Physics, University of Sofia, St. Kliment Ohridski, BG-1164 Sofia, Bulgaria}

\date{}

\begin{abstract}
We have characterized in depth the time response of three detectors equipped with cylindrical LaBr$_{3}$(Ce) crystals
with dimensions of 1-in. in height and 1-in. in diameter, and having nominal Ce doping concentration of 5\%, 8\% and 10\%.
Measurements were performed at $^{60}$Co and $^{22}$Na $\gamma$-ray energies against a fast BaF$_{2}$ reference
detector. The time resolution was optimized by the choice of the photomultiplier bias voltage and the fine tuning of 
the parameters of the constant fraction discriminator, namely the zero-crossing and the external delay.
We report here on the optimal time resolution of the three crystals. It is observed that timing properties are
influenced by the amount of Ce doping and the crystal homogeneity. 
For the crystal with 8\% of Ce doping the use of the ORTEC 935 CFD at very shorts delays in addition
to the Hamamatsu R9779 PMT has made it possible to improve the LaBr$_{3}$(Ce) time resolution from the best 
literature value at $^{60}$Co photon energies to below 100 ps. 
\end{abstract}

\begin{keyword}
fast timing \sep inorganic scintillators \sep LaBr$_{3}$(Ce) \sep Ce concentration
\sep time resolution \sep Hamamatsu R9779 \sep constant fraction discriminator \sep time walk 
\end{keyword}

\end{frontmatter}



\section{Introduction}

\label{intro}

High-density scintillators that exhibit fast response and good energy
resolution are the preferred crystals in a number of applications such as gamma-ray spectroscopy, medical diagnosis
and timing measurements. One of the fastest scintillators commercially
available nowadays is LaBr$_{3}$(Ce) \cite{Loef2002,Shah2003,Gobain}. It has very good timing
properties in addition to high \textgreek{g}-ray efficiency and good
stopping power. Its energy resolution at 662 keV is measured to be as good as
2.8\%, and its photon yield corresponds to 63 photons/keV \cite{Gobain}, which is
much higher than that of other inorganic scintillators \cite{Dorenbos2002208}. 

Due to their superb properties, LaBr$_{3}$(Ce)-based detectors have been the option of choice since 2005 \cite{White2007}
for the application of the Advanced Time-Delayed (ATD) method, which was introduced as a HPGe-gated $\beta-\gamma$ 
electronic timing technique with ultra fast scintillators \cite{Mach1989}.  
The ATD method allows the measurement
of nuclear level lifetimes down to a few picoseconds by using the fast coincidences between
the radiation populating and de-exciting a nuclear energy level \cite{Mach1989,Moszynski1989}.
Nuclear level lifetimes are decisive observables in nuclear physics since
they provide access to reduced transition probabilities between nuclear states, and
therefore insight into the nuclear structure.
Currently \labr\ crystals have become the standard detectors for 
\textit{fast-timing} spectroscopy \cite{Marginean2010,Alhharbi2013,Olaizola2013,Regis2014}.

The sensibility of the ATD method is directly determined by the time resolution of detectors in use, and therefore
the best choice of crystal type (\labr\ in this case), size, shape and doping is required, together with 
the right selection of the coupling photosensor \cite{Fraile2011} is required. The optimization of the electronic circuit and
the detector operation parameters is also needed to achieve the maximum performance \cite{Fraile2013,Fraile201327}
of the set-up.

In this paper we report on the time response of three LaBr$_{3}$(Ce) detectors equipped with 1-in. cylindrical crystals.
The selected photomultiplier tube (PMT) is the Hamamatsu R9779, which is the most performing model available
nowadays in the market \cite{Fraile2011}.
The detector time response has been optimized by fine-tuning of the electronics parameters. 
Three crystals with different nominal Ce concentration were used \cite{Aris2014}, making it possible to assess the effect of
the amount of doping on the time response. 
This work is part of a broader study of the characteristics of ultrafast scintillators detectors, equipped with crystals of
different types, sizes and geometries and coupled to state-of-the-art photomultipliers. 
The aim is to construct a high performance fast-timing array of \labr\ detectors for $\gamma-\gamma$ and $\beta-\gamma$
spectroscopy, and in particular the optimal FAst TIMing array ($FATIMA$) \cite{TDR2015}, which will be placed at the focal plane
of the SuperFRS at FAIR. This instrument belongs to the HISPEC-DESPEC experiment \cite{DESPEC,HISPEC-DESPEC} of the 
Nuclear Structure Astrophysics and Reactions (NUSTAR) collaboration, one of the four pillars of the FAIR project \cite{FAIR}. 

\section{Detector characteristics}

\subsection{The LaBr$_{3}$(Ce) detectors}

Three LaBr$_{3}$(Ce) crystals with identical shapes were studied in order to
characterize their timing properties. The crystals 
were produced by different manufacturers with distinct Ce dopant concentration, but having the same
cylindrical shape with nominal size of 1-in. in height and 1-in. in diameter, and hermetically sealed inside an aluminum housing.
The housing had a thin aluminium window at the entrance, and was fitted with a glass light guide at the coupling side to the photosensor. Inside
the case several layers of light reflector and shock absorbing material assure the stability of the crystal and minimize photon losses.

Crystals were labeled as $A$, $B$ and $C$. Crystal $C$ was the first one produced; it
was grown in 2006 as a test crystal. Since it had been reported that the time resolution
of LaBr$_3$(Ce) crystals improves with the amount of Ce doping \cite{Glodo2005}, Crystal $C$
was made with enhanced Ce doping concentration of 10\%, while standard crystals were commercially
available with 5$\%$ doping. Crystal $A$ was produced in the same year as a commercial crystal using the standard 
crystal growth and production techniques, with a 5\% of Ce. 
Finally, at the end of 2012, Crystal $B$ was produced with an 8\% Ce doping, with the idea of finding a good balance between increased doping
and homogeneity for medium-sized crystals. 

\begin{figure}[H]
\includegraphics[bb=2.30cm 1.2cm 725bp 505bp,clip,scale=0.33]{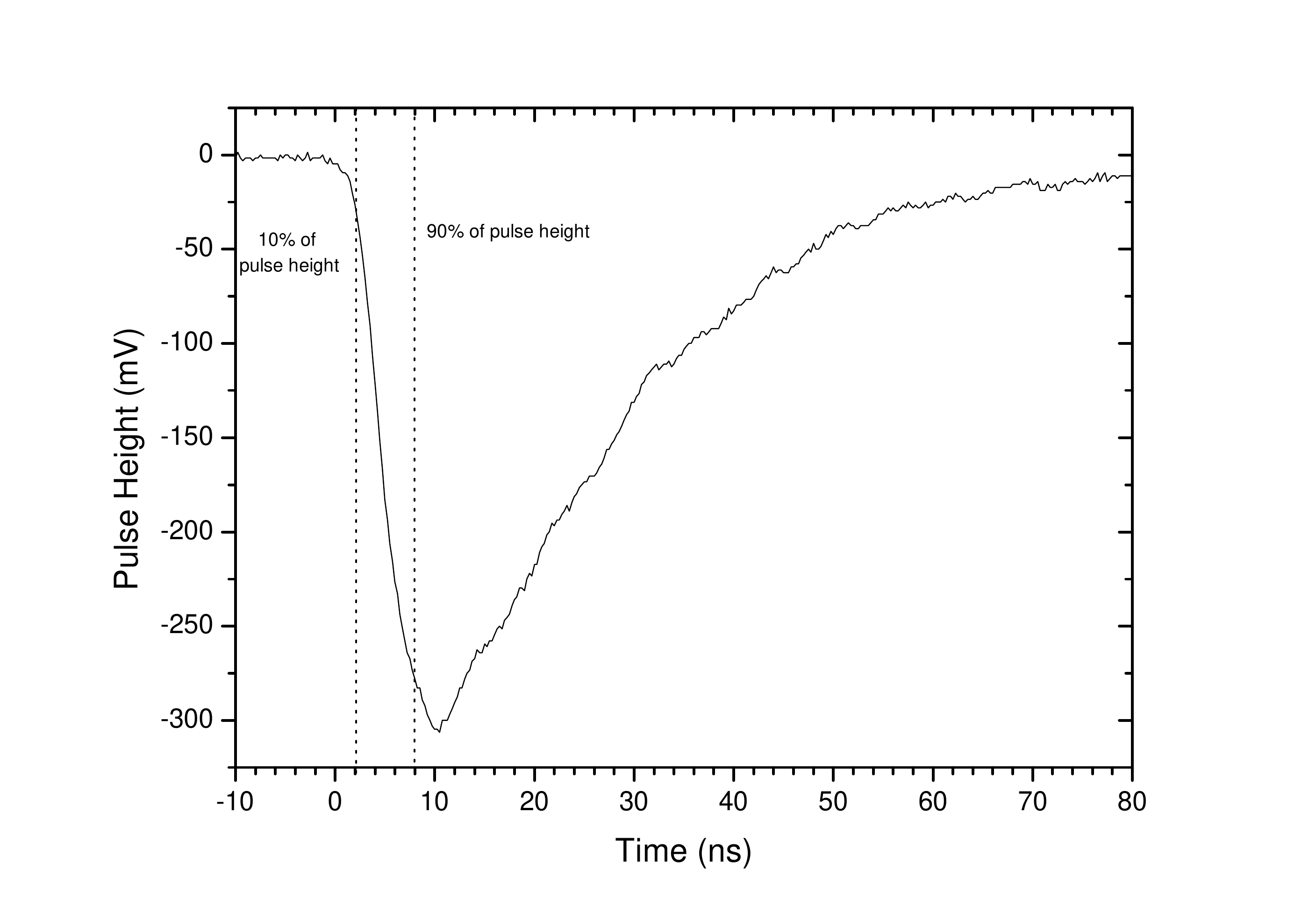}
\caption{\label{fig:Anode-pulse} Anode pulse of Crystal $A${+}Hamamatsu
R9779 acquired with 2 Gsa/s 1 GHz oscilloscope. The used source is a standard $^{137}$Cs and the PMT bias voltage \textminus{}1000 V.
The anode rise time taken from 10\% to 90\% of the maximum pulse height is $\sim$6 ns.}
\end{figure}

The chosen photomultiplier tube fitting the crystal properties is the Hamamatsu
R9779. This tube is optimized for fast timing applications and shows
good timing and energy characteristics \cite{Fraile2011,Fraile2013}.
It is an 8-stage device with a window of 2 in. in diameter, equipped with bialkali photocathode.
It has a typical transit time of 20 ns with transit time spread (TTS) of 250 ps, and anode rise time of 1.8 ns \cite{Hamamatsu2009}. To favour light transmission, crystals were coupled to the PMT by Viscasil
silicon grease and wrapped into opaque tape. 
An example of the anode pulse is plotted in Figure \ref{fig:Anode-pulse}.
PMT high voltages in the range of \textminus{}1200 V to \textminus{}1300 V were found 
to be optimal to provide fast-timing response 
while preserving good energy resolution and linearity as presented in the next sections. 

\subsection{Energy resolution, linearity and efficiency}

In addition to the time response, energy resolution and linearity are also important characteristics to be considered. 
A non-linear energy response implies high-order polynomial energy corrections, while bad energy resolution could lead to the 
inability of precise \textgreek{g}-ray energy determination and hamper transition selection. 

\begin{figure}[h]
\begin{raggedright}
\includegraphics[bb=2.75cm 1.cm 735bp 510bp,clip,scale=0.34]{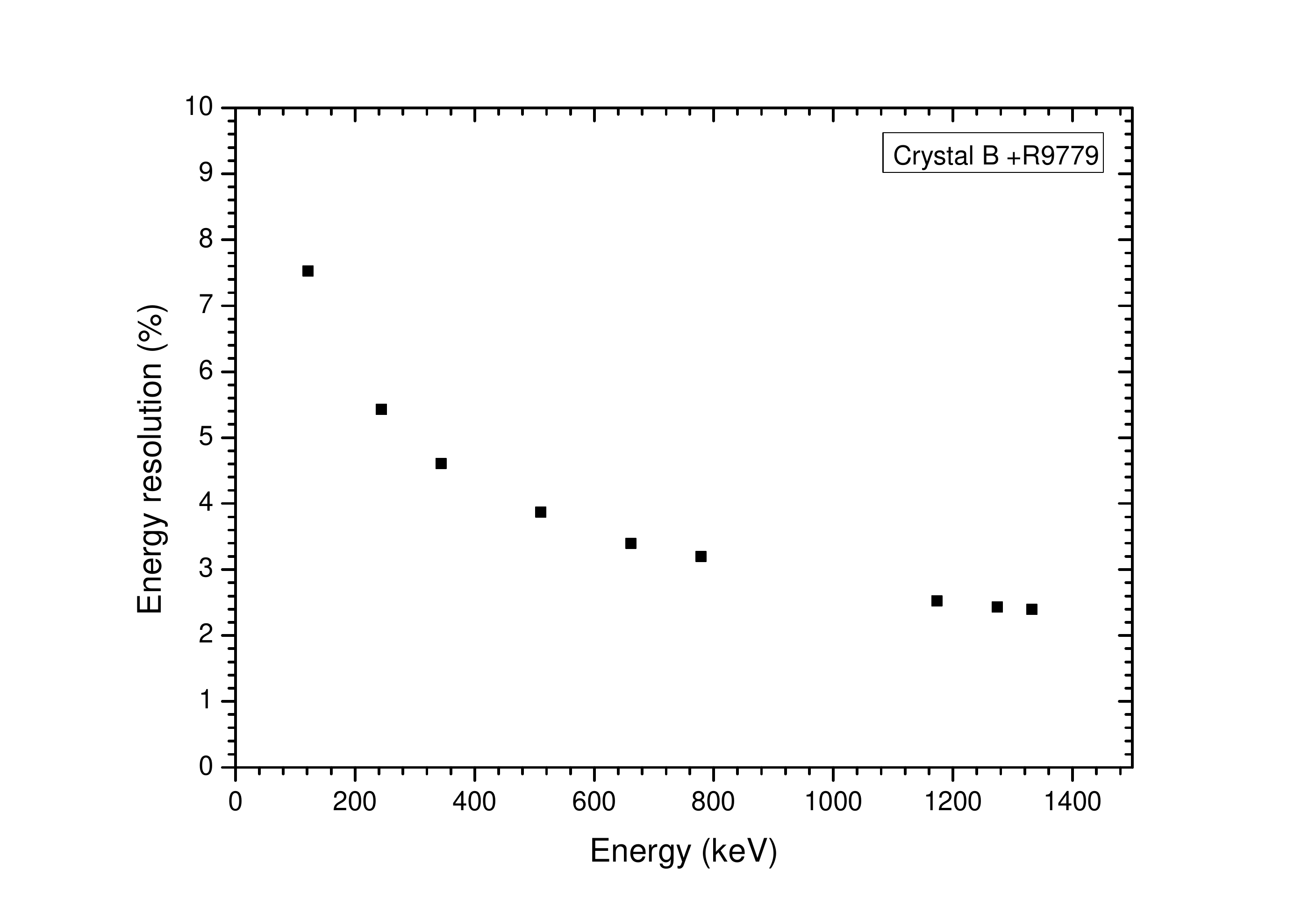}
\par\end{raggedright}

\caption{\label{fig:Eresolution vs E} Energy resolution of Crystal $B${+}Hamamatsu R9779 as a function
of the energy. The PMT is operated at \textminus{}1300 V.}
\end{figure}

We report here the energy resolution of Crystals $A$, $B$ and $C$ at 662 keV ($^{137}$Cs), given as the
ratio between the FWHM of the \textgreek{g}-ray full energy peak and its energy, corrected for non-linearity \cite{Fraile2013}. 
The resolution was measured at the PMT bias voltages that provide the best time resolution.
The relative energy resolution of Crystals $A$, $B$ and $C$, once corrected for non-linearity, is measured to be 3.4\%, 3.4\% and 4.0\% 
respectively. These values are worse than the number quoted by the
manufacturer, 2.8\% (at 662 keV), due to the use of the fast R9779 PMT \cite{Hamamatsu2009}, which is optimized for timing measurements, and provides the best timing performance at the expense of slightly worse energy resolution. To quantify this effect Crystal $A$ was also tested with a second PMT, designed for energy measurements (Hamamatsu R6231), yielding 2.9\% (at 662 keV), value that matches the specifications. Table \ref{tab:Summary-table-of} summarizes the energy resolution measured for the three crystals with both PMT models (R9779 and R6231) and Figure \ref{fig:Eresolution vs E} illustrates Crystal~$B$ energy resolution as a function of the energy in the range from 122 to 1332 keV.

\begin{table}[H]
\begin{centering}
{\small{}}
\begin{tabular}{|c|c|c|c|c|}
\hline 
{\small{Crystal}} & {\small{Ce (\%)}} & {\small{PMT}} & {\small{HV(V)}} & {\small{Er (\%)}}\tabularnewline
\hline 
\hline 
{\small{A }} & {\small{5}} & {\small{R9779 }} & {\small{1300}} & {\small{3.4}}\tabularnewline
\hline 
{\small{A }} & {\small{5}} & {\small{R6231}} & {\small{1000}} & {\small{2.9}}\tabularnewline
\hline 
{\small{B }} & {\small{8}} & {\small{R9779 }} &{\small{1300}} & {\small{3.4}}\tabularnewline
\hline 
{\small{C }} & {\small{10}} & {\small{R9779 }} & {\small{1200}} & {\small{4.0}}\tabularnewline
\hline 
\end{tabular}
\par\end{centering}{\small \par}

\caption{\label{tab:Summary-table-of} Relative energy resolution Er of the three crystals (A, B and C)
at $^{137}$Cs energy (662 keV). The uncertainty in the values is 0.1\%.
The crystals were coupled to Hamamatsu R9779 and R6231 PMTs.}
\end{table}

The dependence of the energy resolution on the Ce concentration was assessed as well. 
It has been shown in \cite{Glodo2005} that up to 5\% of Ce concentration the 
LaBr$_{3}$(Ce) photon yield increases \cite{Glodo2005} and hence the energy resolution also improves. 
However, a further Ce increase leads to
constant or even lower energy resolution \cite{Drozdowski2008}. In our case, Crystals $A$ (5\% of Ce) and $B$ (8\% of Ce)
have the same energy resolution, while Crystal $C$, with a 10\% nominal Ce concentration, shows the worst
value among the three, 4.0\%. The measured photon yield for crystal $C$ is $\sim$8\% lower than for crystal $A$, 
in accordance with \cite{Glodo2005}, but also the photon yield for crystal $B$ with 8\% Ce doping is higher than
for crystal $A$ by $\sim$10\%. 
Given that the contributions to the intrinsic resolution are the non proportional 
response (which is low for \labr\ \cite{Shah2003}) and the inhomogeneities, which cause local variations in the scintillation light output \cite{Dorenbos1995}, the 
main reason for the worse energy resolution for the highly doped crystal $C$ could be 
explained by Ce inhomogeneities in the crystal.

\begin{figure}[h]
\includegraphics[bb=2.50cm 1cm 740bp 520bp,clip,scale=0.34]{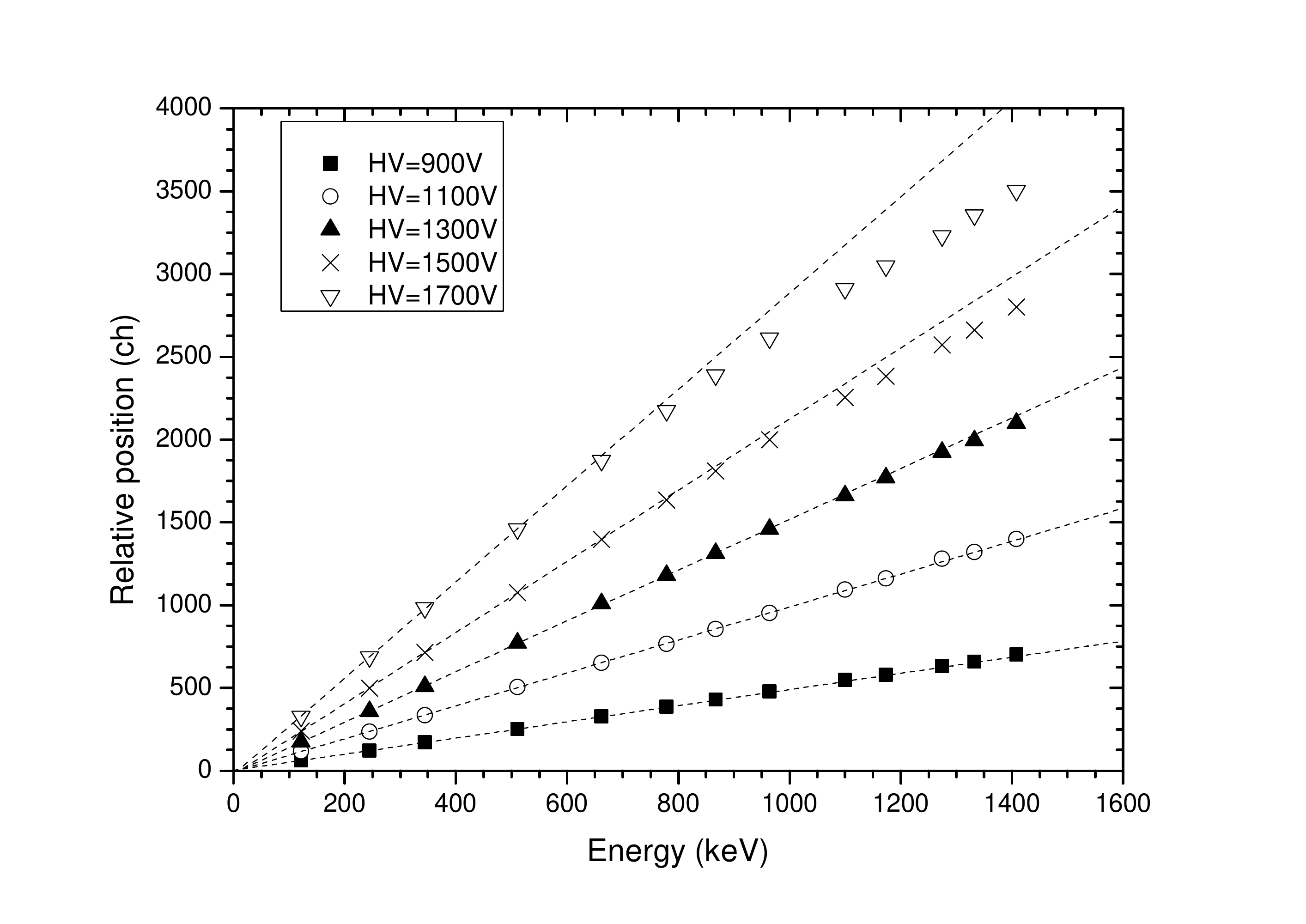}

\caption{\label{fig:Linearity} Functional relation between the peak position and
the actual $\gamma$ energy taken at five different HV values, for Crystal $B$ coupled to the R9779 PMT. 
Only the first 4 points of each data set were linearly fitted.}
\end{figure}

Detectors based on high photon yield scintillators such as LaBr$_{3}$(Ce) may display
non-linear behaviour, even with good light yield proportionality.
This may be caused by space-charge effects in the photomultiplier tube \cite{Mos2006}.
In order to check the energy linearity of our detectors, we have measured the functional relationship between
the peak position (signal amplitude) and the real \textgreek{g}-ray
energy from a $^{152}$Eu source for the three crystals at bias voltages in the range from \textminus{}900 V
to \textminus{}1700 V.
We have found that LaBr$_{3}$(Ce) detectors behave linearly over the
analysed energy range, and especially at PMT voltage values from \textminus{}900 V
to \textminus{}1300 V. Small deviations from linearity appear at \textminus{}1300 V,
where the best time resolution is achieved. At higher voltages than \textminus{}1300 V the behaviour deviates significantly from linearity.
It has to be noted that the three detectors have the same response, supporting the idea that the main influence 
comes from the PMT gain variance \cite{Dorenbos1995}.
As an example, Figure \ref{fig:Linearity} shows the signal amplitude (peak position) versus the \textgreek{g}-ray energy for Crystal $B$ 
at different PMT voltages. To illustrate the departure from the linear behaviour only the first four points of each 
data set were linearly fitted, and the fit was extrapolated up to the 1600 keV range.

It should be underlined that at the operational voltage giving the best timing performance (see subsection \ref{sec:timing}), both the energy resolution 
and linearity of the \labr$+$R9779 detectors are also optimal, and thus well suited for spectroscopy measurements. 

\begin{figure}[h]
\includegraphics[bb=2.3cm 1cm 750bp 530bp,clip,scale=0.33]{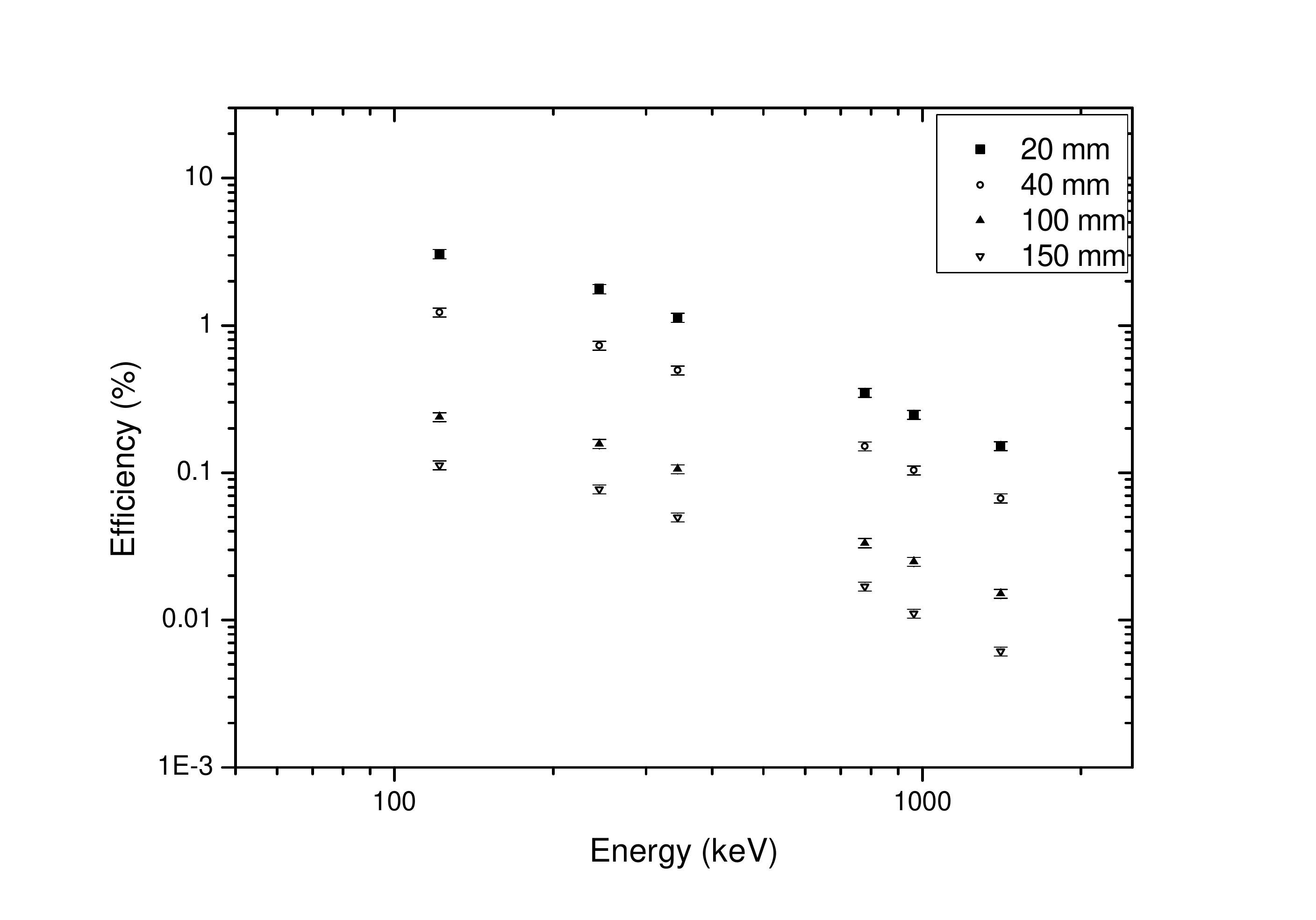}

\caption{\label{fig:=0003B3-ray-detection-effciciency}Full energy peak detection
efficiency of Crystal B-R9779 operated at \textminus{}1300 V. The $\gamma$-ray efficiency
was measured at 20, 40, 100 and 150 mm from the crystal end-cup in the energy range from 122 to 1408 keV.}
\end{figure}

Another interesting feature for the design of a fast-timing array based on this kind of detectors
is the \textgreek{g}-ray detection efficiency. In the present study the absolute full energy peak
efficiency was measured with a $^{152}$Eu source placed at several distances from the crystal housing
end-cap. Results for Crystal B at the distances of 20, 40, 100 and 150 mm
are shown in Figure \ref{fig:=0003B3-ray-detection-effciciency}. The crystal was 
coupled to the R9779 PMT, which was powered to 1300 V. 

\section{Description of the measurements}
\subsection{Experimental set-up for time-resolution measurements}

The time characterization of each LaBr$_{3}$(Ce) detector was performed
by coincidence measurements against a BaF$_{2}$
reference detector, whose time response was earlier determined by the use
of three identical BaF$_{2}$ detectors with equal response \cite{Fraile2013}.
Both detectors, the reference BaF$_{2}$ and the one under study,
are fixed to a frame, aligned and held in close geometry, with the radioactive
source placed in between. The small BaF$_{2}$ reference crystal was
coupled to a Photonis XP2020-URQ PMT operated at \textminus{}2300
V bias, and the three LaBr$_{3}$(Ce) crystals described above were coupled to the aforementioned
Hamamatsu R9779 PMT. The negative anode signals from both PMTs are used for
timing measurements, directly fed to an ORTEC 935 Constant Fraction Discrimination (CFD), where
the timing of every signal is determined by the constant
fraction or by the leading edge methods, generating an appropriate fast
negative output pulses. After the ORTEC 935 both CFD outputs are
sent to an ORTEC 567 Time to Amplitude Converter (TAC), which works
as an time comparator, and provides square pulses whose
height corresponds to the time difference between the start (\baf) and the
stop (\labr) signals. 

The last dynode positive signals are used for the energy measurement. They 
are processed firstly by an ORTEC 113 preamplifier
and then by a TENNELEC TC 247 spectroscopic amplifier module. 

The TAC amplitude signal and the two energy signals after the amplifier are fed to ADCs.
For each valid TAC event a gate signal is generated by means of a Gate and Delay Generator
enabling the acquisition and digitization in the three ADCs.
Coincidence list-mode data are stored for further analysis. 

\subsection{Timing measurements}

Timing data were collected for $^{60}$Co energies and for $^{22}$Na (511 keV). 
Data were processed and analysed with the SORTM
software \cite{Mach2011}. 
Since the readout electronics are very temperature sensitive, the first step of the analysis
consists in the identification of shifts in the time spectrum centroids and subsequent 
correction. Owing to temperature stabilization, only very small variations well within
the uncertainty of the centroid position, are observed. 
The second step consists in the offline energy gate selection on full energy peaks (FEP)
at each detector, in order to project to the time spectrum for the selected energies. 
For the $^{60}$Co source, gates are set at the 1173 and 1332-keV peaks. Thus, there are two time spectra
sorted out, one for each of the two possible combinations: when the 1173-keV FEP is selected in the \baf\ and
the coincident 1332-keV FEP is detected by the \labr\ detector, and when
the reversed combination is chosen. The time resolution for $^{60}$Co
is reported here for the summed spectrum. For the $^{22}$Na source
narrow gates are set at the full width at half maximum (FWHM) of the 511-keV peak, generating only
one time spectrum. All steps of the sorting procedure are explained
in more detail in \cite{Fraile2013,Mach2011}.

Once the time spectrum is generated, the coincidence resolution time (CRT) is obtained by measuring 
the FWHM of the time peak, and the individual \labr\ detector time resolution is extracted by
de-convoluting the \baf\ contribution, 81\textpm{}2 ps at $^{60}$Co energies
and 120\textpm{}2 ps at 511 keV ($^{22}$Na).
The reported values in Table \ref{tab:Best-FWHM} refer to the individual \labr\
FWHM time resolution.

\subsection{Optimization of the time resolution}\label{sec:timing}

The first key elements influencing the detector timing performance is the 
PMT itself. As discussed above, the Hamamatsu R9779 was chosen in the present work, given that it has shown much superior
features than other fast PMTs also optimized for high photon yield scintillators.

The next important element is the constant fraction discriminator. 
In this study we used the ORTEC 935 CFD due to its proven capability of working
with very fast scintillators providing excellent time resolution and time walk below \textpm{}50 ps over
a 100:1 dynamic range \cite{ORTEC935}. 
To cope with the very short pulses and provide
optimal time resolution, the ORTEC 935 CFD is equipped with an especial transformer
(XFMR) for constant-fraction shaping and a selectable internal delay
of \textminus{}1 ns (W1). 
The total delay selected at the ORTEC 935 CFD determines the time of 
the incoming pulses, defined by the zero-crossing of the bipolar
signal with the baseline. The bipolar signal originates from the addition of the 
attenuated signal by a factor of 20\% and delayed \textminus{}1
ns when W1 jumper is removed, and the original signal retarded by the
CFD external delay, which is implemented as a LEMO cable at the front panel.
The use in this work of very short total delays of 0.6 ns or shorter has made it possible
to obtain time resolution values never achieved before.

\begin{figure}
\begin{tabular}{|c|}
\hline 
\includegraphics[bb=0.7cm 30bp 695bp 510bp,clip,scale=0.3]{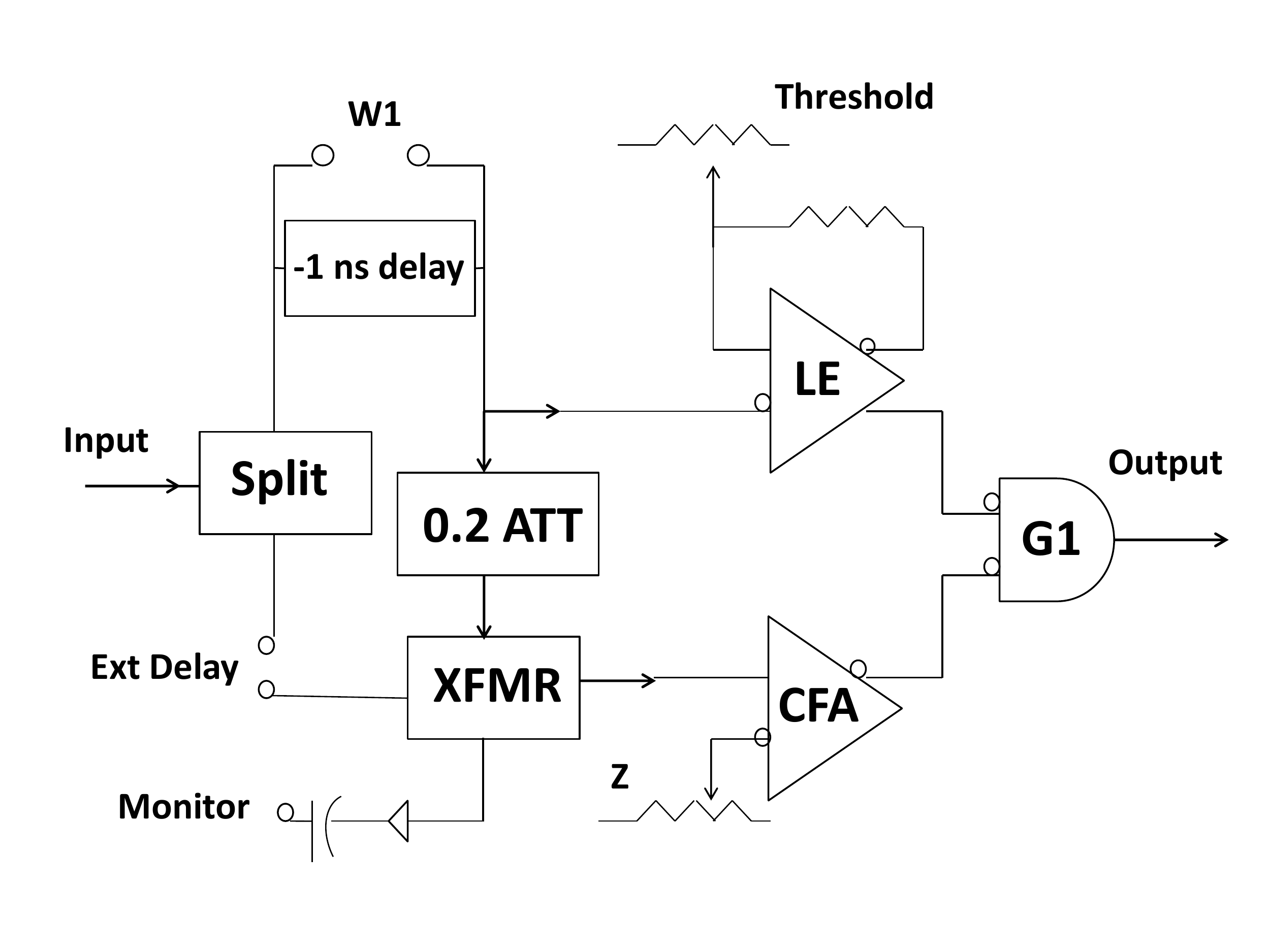}\tabularnewline
\hline 
\end{tabular}
\caption{\label{fig:CFD diagram}Simplified block diagram illustrating the operation
of the ORTEC 935 CFD, adopted from \cite{ORTEC935}}
\end{figure}

Figure \ref{fig:CFD diagram} sketches the operation of the ORTEC 935 module.
When the external delay is shorter than 1 ns, there is no bipolar
output from the XFMR and the unit works as a leading edge (LE) comparator,
triggering the incoming pulses when they go above the adjusted threshold, as 
already discussed in \cite{Szczesniak2007}.
Time walk, drifts and time jitter are the most common problems that deteriorate
time resolution when a leading edge discriminator
is used. However, when the pulse rise time is sharp enough and with low
jitter, this timing triggering method can result very efficient. This issue will be explored
for our 1 in. $\times$ 1 in. crystals by the operation of the ORTEC 935 in either LE or CFD mode.

In our study the time resolution was optimized through an iterative procedure, starting by the 
the optimization of the external CFD delay, followed by the fine tuning of the
zero crossing Z, and finally of the PMT bias voltage.
During the measurements the internal delay W1 jumper on the ORTEC 935 CFD was removed,
setting an internal delay of \textminus{}1.0 ns. We have then systematically explored
a wide range of external delays from 0.5 to 20 ns. The starting value for such optimization
of the external delay is the measured anode rise time at 662 keV, as depicted in Figure \ref{fig:Anode-pulse}.

Care must be taken when very short external delays like 1.5 ns
are used, as the time walk becomes very sensitive to the Z value. 
The time resolution does not significantly change when Z
is varied, but important modifications may happen on the time walk. 
Due to this fact, the zero crossing Z at the
ORTEC 935 module was also careful investigated, firstly by observing the signals at the scope to define a
valid operational range, and secondly by fine tuning the Z parameter within this range. The aim was to obtain a smooth time walk curve without compromise of the time resolution.

The final parameter to be adjusted is the PMT bias voltage, which 
influences the electron multiplication and collection, and secondary emissions inside the
PMT. The time resolution is not as sensitive to this parameter as it is
to the external delay, but it is relevant regarding the linearity
and energy resolution. Space-charge effects may occur, leading to non-linear
performance or affecting the energy resolution. The initial
values were chosen to obtain a 1 V anode amplitude for 1 MeV incident $\gamma$-rays in the crystal. 
The best HV value is selected by taking into
account three factors: linearity, energy resolution and time resolution.

\section{Results and discussion}

\subsection{Dependence on the CFD external delay}

\begin{figure}[t]
\includegraphics[bb=2.4cm 1.2cm 740bp 530bp,clip,scale=0.33]{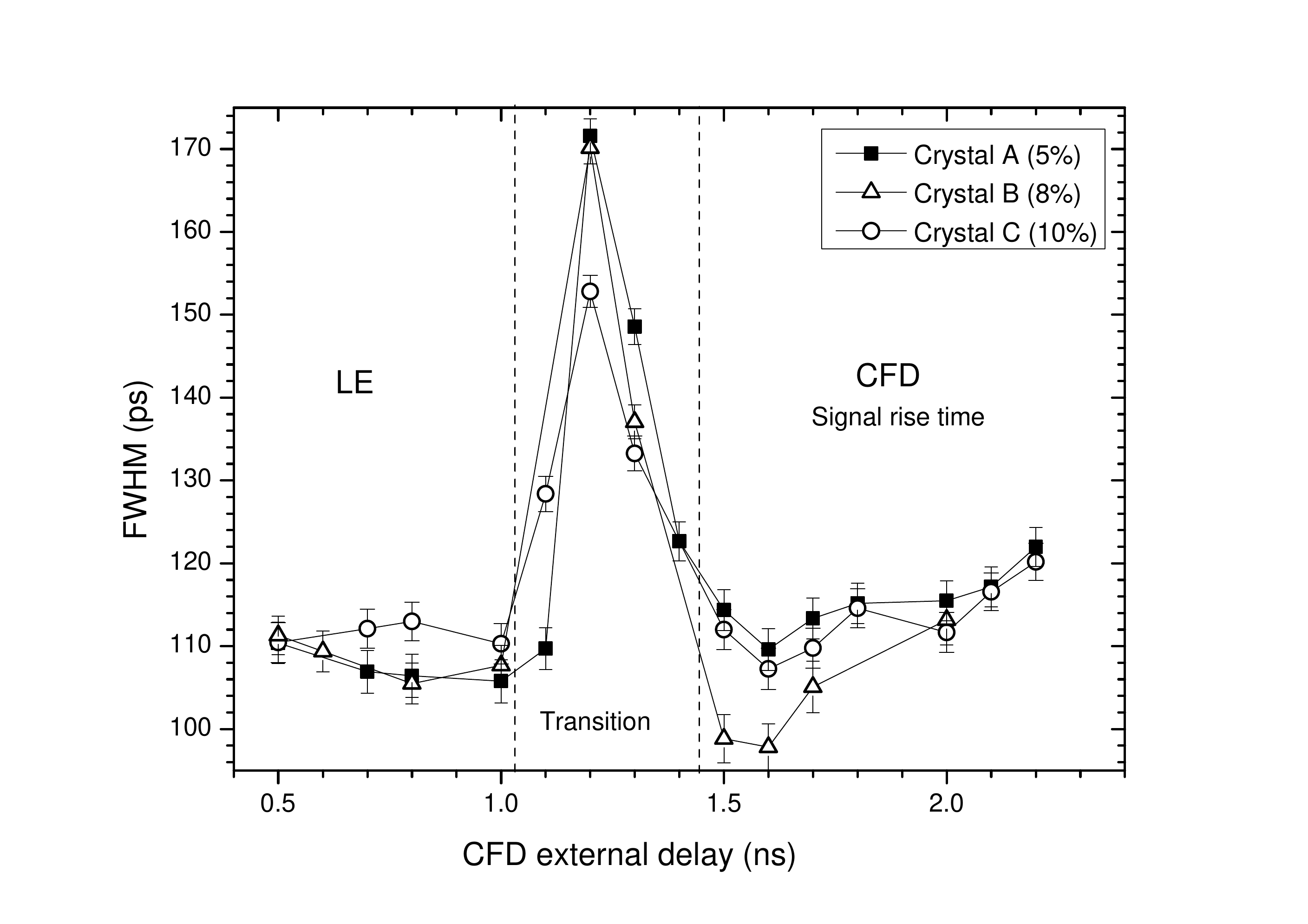} \\
\includegraphics[bb=2.4cm 1.2cm 740bp 530bp,clip,scale=0.33]{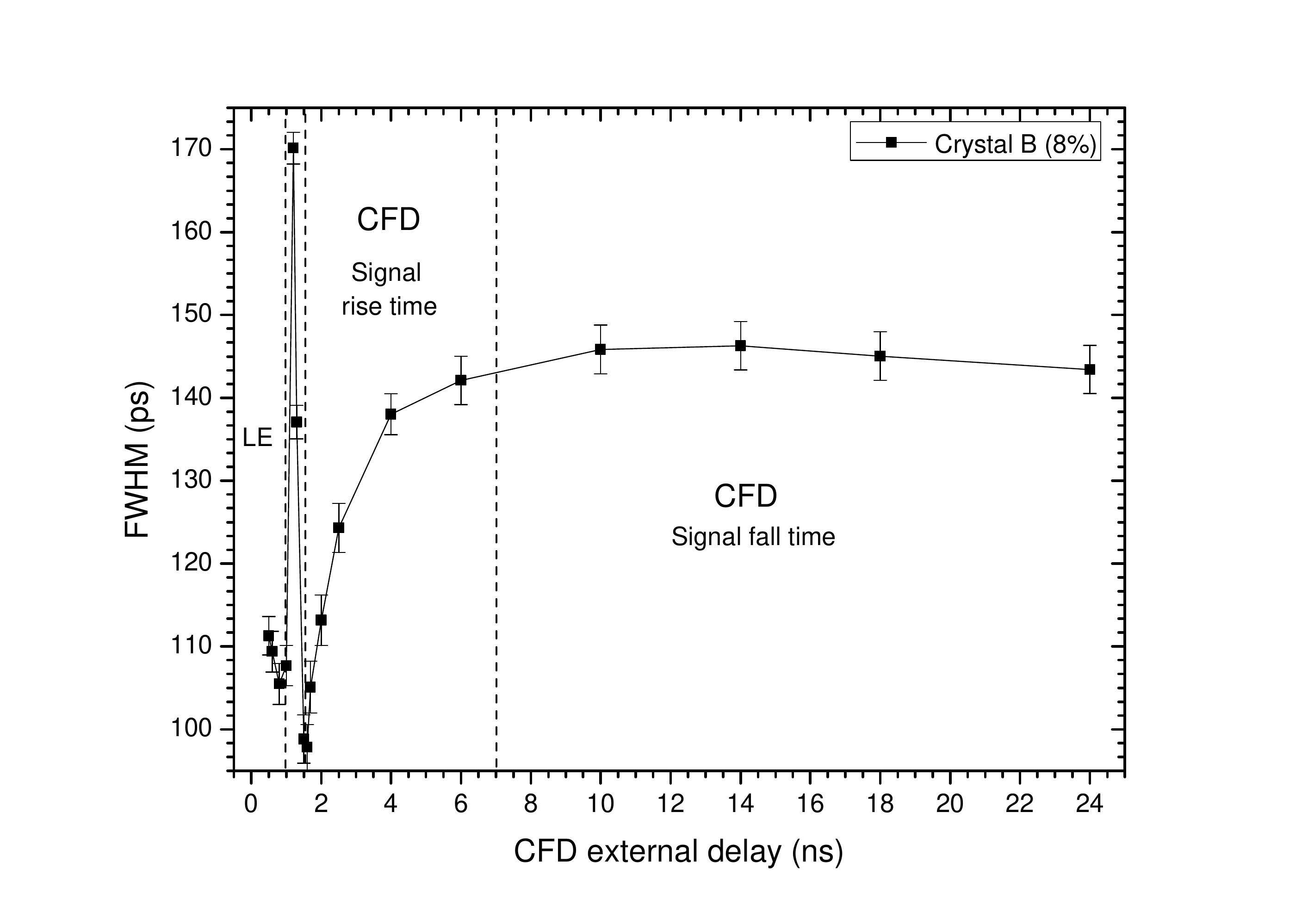}\caption{\label{fig:FWHM-delay}Time resolution of Crystals $A$, $B$ and $C$ as a function of the external delay, for short values (top
panel). The bottom panel shows the FWHM dependence for detector $B$ for the whole range of study.
Detectors $A$ and $B$ are operated at $-$1300 V and $C$ at $-$1200 V. The CFD settings
correspond to Z=2.0 mV and Th=$-$508 mV.}
\end{figure}

The time resolution of Crystals $A$, $B$ and $C$ coupled to the R9779 PMT is plotted in Figure \ref{fig:FWHM-delay} for $^{60}$Co energies,
as a function of the ORTEC 935 CFD external delay.
The best time resolutions are obtained for an external delay of 1.6 ns. For Crystal $B$ (8\% Ce doping) this value 
is as good as 98\textpm{}2 ps. The best values deteriorate by more than 50\% when the external delay is modified.
The upper plot on Figure \ref{fig:FWHM-delay} illustrates the FWHM versus external
delay dependence in the short delay range for Crystals $A$, $B$ and $C$. The
bottom panel displays the FWHM as a function of the CFD external delay for Crystal B in the entire range.
The trend is identical for all three detectors.

There are four distinct regions of operation of the CFD as a function of the external delay. 

When the external delay is smaller than 1 ns (0.5$-$0.8 ns) the 935 unit triggers as a LE discriminator providing
a good time resolution independently of the external delay and
comparable to that obtained by the CFD triggering procedure. This
is possible because the incoming pulses have a sharp rise time and show a low level of 
jitter. The leading edge triggering fractions were also investigated here in order 
to achieve the best values of FWHM time resolution. 
The optimal leading edge triggering level is reached when the threshold control at the ORTEC 935 
is set at $-$508 mV, equivalent to a real input voltage level of 64 mV. At this operation bias, this 
corresponds to a 3.75{\%} of the anode pulse height for $\gamma$-rays of 1332 keV.

Secondly, there is an unstable region between 0.9 and 1.3 ns where the
time resolution severely increases, as a consequence of the regime
transition. The operation 
in this mode should be avoided. 

Thirdly, for external delays higher than 1.4 ns, the ORTEC 935
works correctly as a CFD discriminator. In this regime the time resolution
achieves a minimum at 1.6 ns for the three crystals, and afterwards it smoothly increases
until it reaches a fairly constant value of FWHM for external delays larger
than 7 ns (effective shaping delay equal to the pulse rise time, 6 ns). 

Finally, for long delays the CFD operates on the decay of the time pulses, yielding and almost flat behaviour
of the time resolution, but at the expense of worse values, which reach (146\textpm{}2 ps) at 14 ns for 
Crystal $B$. 

\subsection{Time resolution}

The best time resolutions that we have obtained for Crystals $A$, $B$ and $C$ at $^{60}$Co energies 
are 106\textpm{}2, 98\textpm{}2 and
107\textpm{}2 ps respectively. These values are achieved when very
short external delays around 1.6 ns are set at the ORTEC 935 and the
Hamamatsu R9779 is operated in the range $-$1200 or $-$1300 V. 

We note already at this point that the time resolution is not drastically affected by the PMT bias voltage.
In fact the response is nearly flat. There is only a 
small difference in FWHM, below 5 ps, 
between the best and the worse values in the studied
range, which varies from \textminus{}1100 to \textminus{}1700 V.

\begin{figure}[h]
\includegraphics[bb=2.4cm 1.2cm 740bp 530bp,clip,scale=0.32]{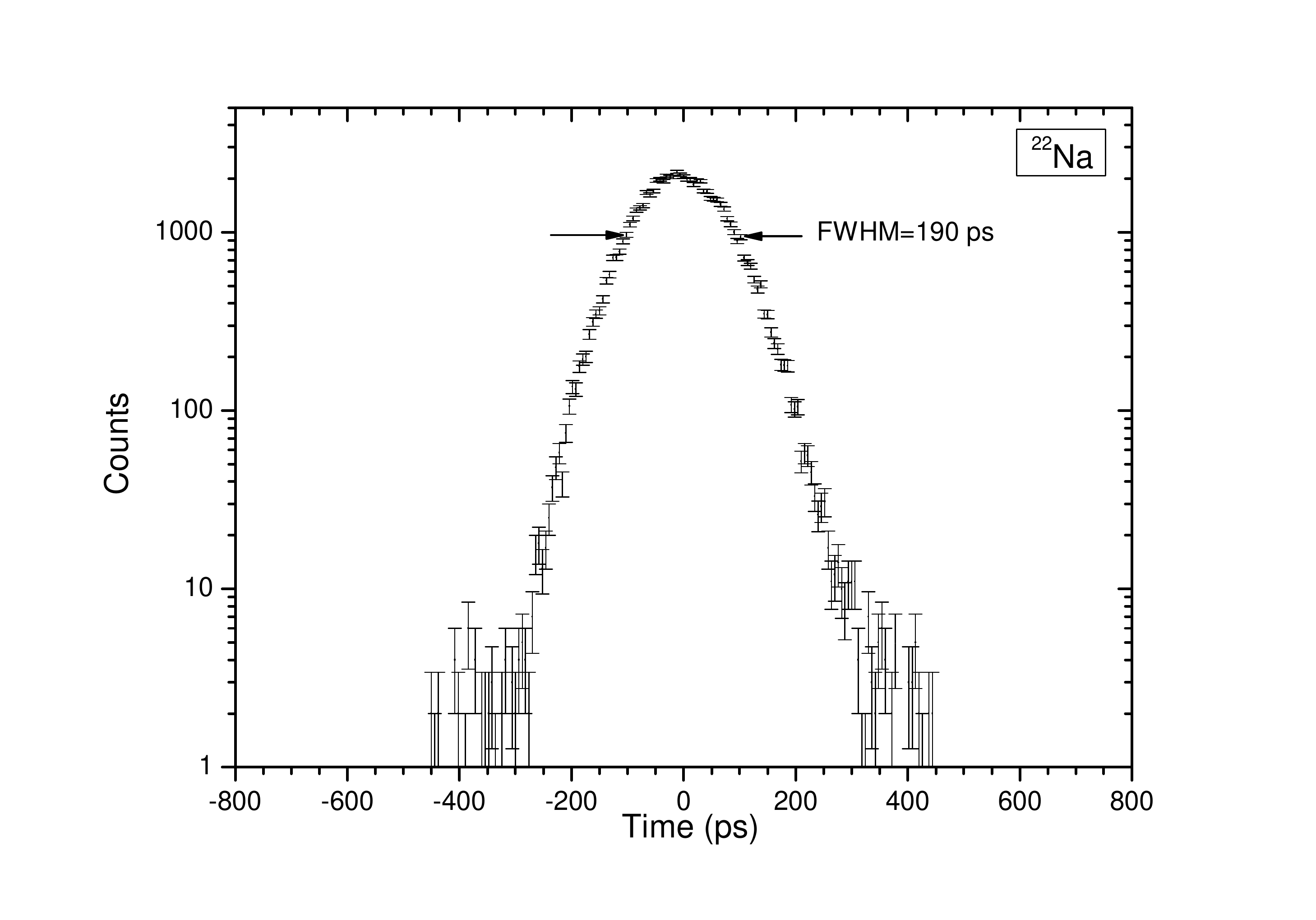} \\
\includegraphics[bb=2.4cm 1.2cm 740bp 530bp,clip,scale=0.332]{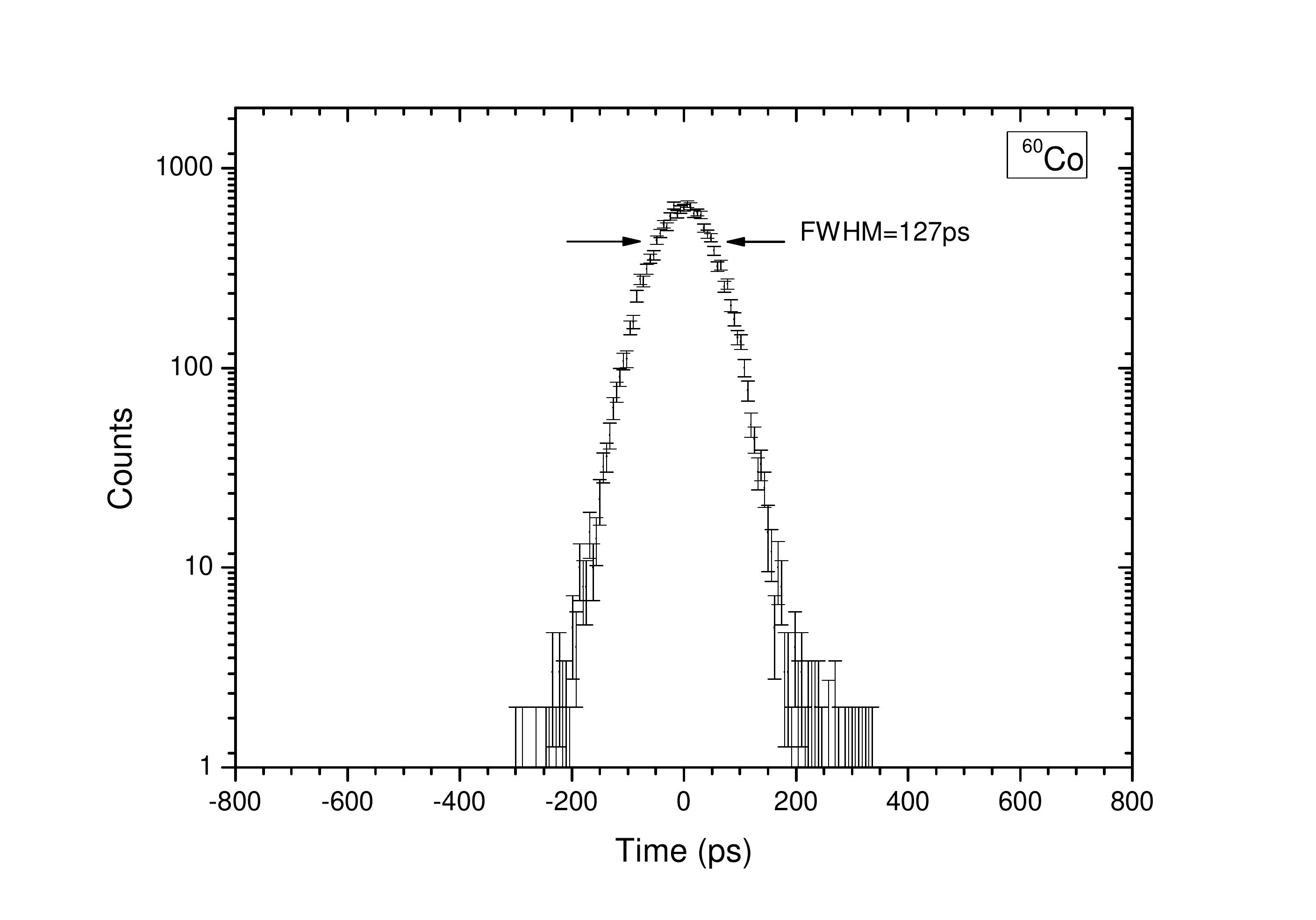}
\caption{\label{fig:FWHM-60Co}Time spectrum for Crystal B\textendash{}R9779
detector against the reference detector measured with a $^{22}$Na
source (top panel). The FWHM resolution of 190 ps is the convolution of the
contributions from the reference detector of 120\textpm{}2 ps and
the LaBr$_{3}$(Ce) unit of 148\textpm{}2 ps. The bottom panel shows the
time spectrum measured for the same detectors using a $^{60}$Co source.
Here the time resolution of 127 ps includes individual contributions of 81\textpm{}2 ps
from the reference detector and  98\textpm{}2 ps from LaBr$_{3}$(Ce). }
\end{figure}

Crystal $B$, which was the last one produced and contains 8\% of Ce dopant,
provides the best value of time resolution, below 100 ps at $^{60}$Co energies. 
The time spectra of Crystal B$+$R9779 is shown in Figure \ref{fig:FWHM-60Co}.
The FWHM resolution of 190 ps using a $^{22}$ Na source includes the individual contributions from
the reference detector of 120\textpm{}2 ps and the LaBr$_{3}$(Ce) unit
of 148\textpm{}2 ps. Similarly the time resolution of 127 ps using a $^{60}$Co source is the convolution of 
the individual reference detector resolution of 81\textpm{}2 ps and that of the LaBr$_{3}$(Ce) detector,
which is 98\textpm{}2 ps. 

The best FWHM values of time resolution at $^{60}$Co energies and 511 keV ($^{22}$Na) for the three
individual \labr\ crystals are summarized in Table \ref{tab:Best-FWHM}.
The best value for Crystal $A$ (106\textpm{}2 ps) is achieved
when the ORTEC 935 triggers as a LE discriminator, due to the very steep pulse rise time (see Figure \ref{fig:Anode-pulse}). 
The value increases up to 110\textpm{}2 ps when the module works as CFD discriminator. 
For the other crystals the values in LE operation
mode are close to the best CFD mode FWHM values, showing that leading edge operation is viable 
whenever the time walk is not relevant.

\begin{table}[H]
\begin{singlespace}
\begin{centering}
{\footnotesize{}}
\begin{tabular}{|c|c|c|c|c|c|}
\hline 
\multirow{2}{*} {\scriptsize{Crystal}} & \scriptsize{External} & \scriptsize{HV} & \scriptsize{ORTEC} & {\scriptsize{FWHM}} & \scriptsize{FWHM} \tabularnewline
 & \scriptsize{Delay (ns)} & \scriptsize{ (V)} & {\scriptsize{ MODE}} & {\scriptsize{ $^{60}$Co (ps)}} & {\scriptsize{ $^{22}$Na (ps)}}\tabularnewline
\hline 
\hline 
\scriptsize{A } & \scriptsize{1.6 } & \scriptsize{1300 } & {\scriptsize{CFD}} & {\scriptsize{110\textpm{}2}} & {\scriptsize{164\textpm{}2}}\tabularnewline
\hline 
{\scriptsize{A }} & {\scriptsize{0.8 }} & {\scriptsize{1300 }} & {\scriptsize{LE}} & {\scriptsize{106\textpm{}2}} & {\scriptsize{158\textpm{}2}}\tabularnewline
\hline 
{\scriptsize{B }} & {\scriptsize{1.6 }} & {\scriptsize{1300 }} & {\scriptsize{CFD}} & {\scriptsize{98\textpm{}2}} & {\scriptsize{148\textpm{}2}}\tabularnewline
\hline 
{\scriptsize{C }} & {\scriptsize{1.6 }} & {\scriptsize{1200 }} & {\scriptsize{CFD}} & {\scriptsize{107\textpm{}2}} & -\tabularnewline
\hline 
\end{tabular}
\par\end{centering}{\footnotesize \par}
\end{singlespace}
\caption{\label{tab:Best-FWHM}Best values of FWHM time resolution for Crystals
A, B and C at $^{60}$Co energies. The zero crossing value (Z) is
set at 2.0 mV and the CFD threshold at -508 mV }
\end{table}

The timing properties of LaBr$_{3}$(Ce) detectors are influenced by the
internal structure of the crystals and the Ce concentration.
In contrast to the energy resolution, the time response is expected to improve when 
Ce is increased above 5\% \cite{Glodo2005}. Crystal $B$, which contains
8\% of dopant, gives a superior time resolution (about 10\% better) than Crystals
$A$ and $C$, even though the nominal doping concentration of crystal $C$ is higher (10\%).
In connection with the influence on the energy resolution discussed above a possible 
explanation may be a non-uniform distribution of the Ce inside the scintillator, 
since it was grown as a test crystal when the production technique was not as developed as it is now. 
Local variations in the scintillation light output may affect the timing response

Not only is the time resolution achieved with Crystal $B$ the best 
among the three detectors studied in the present work, but it also stands as the best result 
reported to date for a standard cylindrical 1-in. LaBr$_{3}$(Ce) crystal \cite{Mos2006}.
This has been possible thanks to the good properties of the Hamamatsu R9779 PMT in conjunction
with the use of the ORTEC 935 CFD at very short delays.
The excellent time resolution of 98\textpm{}2 ps at $^{60}$Co energies is a step forward for the 
application of the ATD method in nuclear spectroscopy. 

\subsection{Zero crossing and time walk}

The time walk optimization by the adjustment of the zero crossing potentiometer in 
the 935 CFD only produces minor changes in the width of the time distribution. The
systematic analysis yields the best time resolution for a zero crossing parameter of Z=2.0 mV. 

\begin{figure}[t]
\includegraphics[bb=2.4cm 1.2cm 740bp 530bp,clip,scale=0.33]{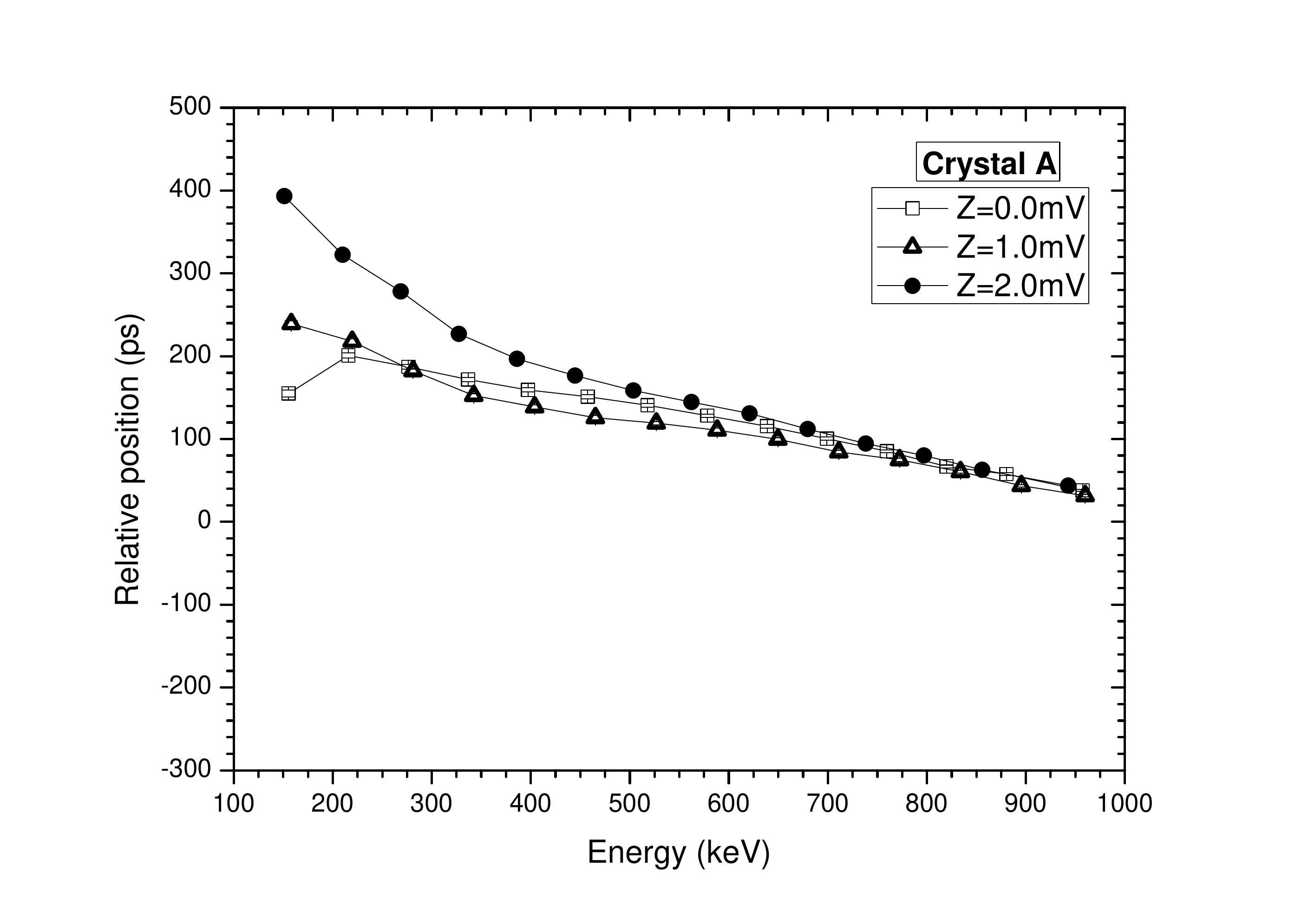} \\
\includegraphics[bb=2.4cm 1.2cm 740bp 530bp,clip,scale=0.34]{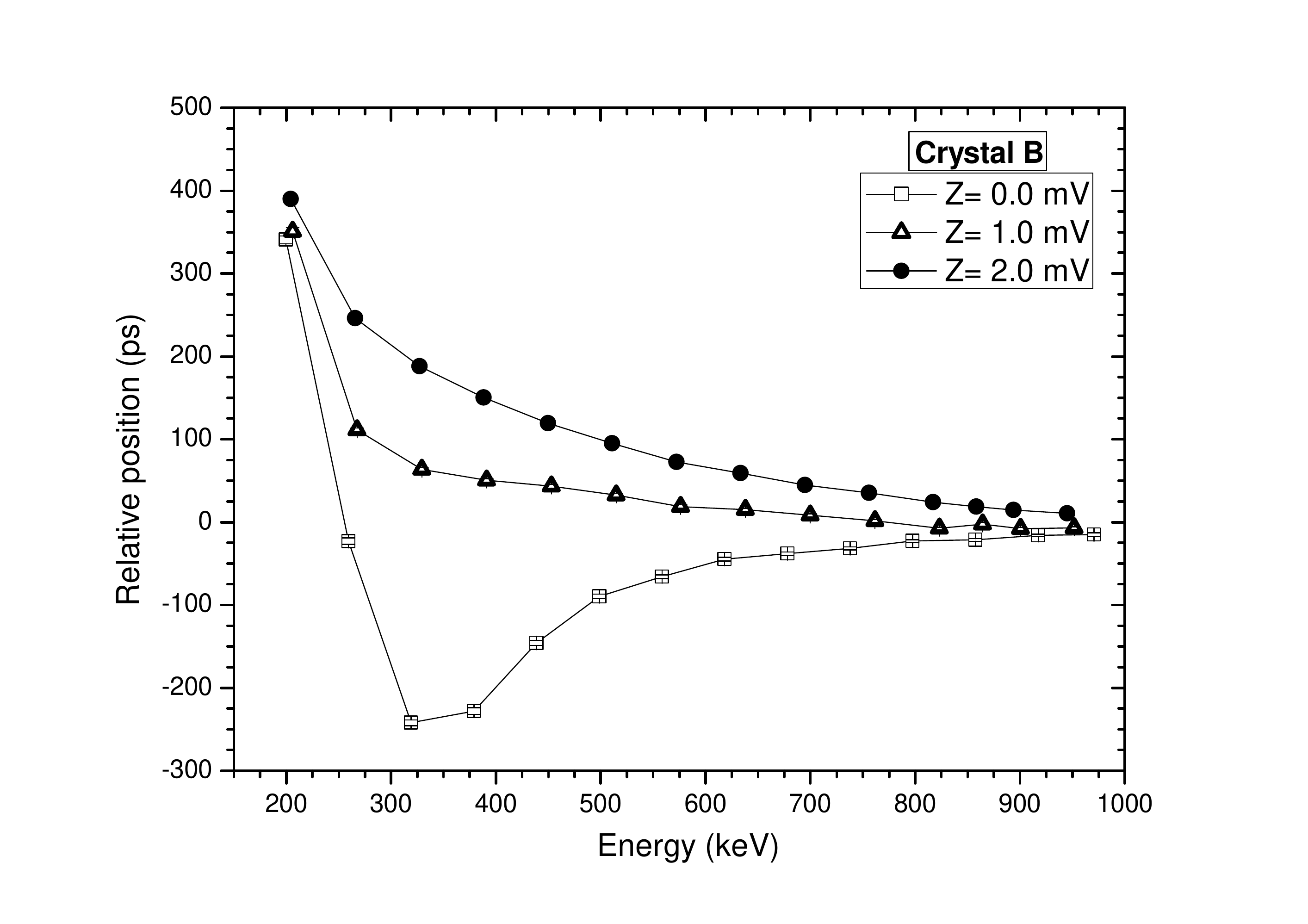}
\caption{\label{fig:walk curve}Compton time walk response of Crystal $A$ (top
panel) and $B$ (bottom panel) coupled to the Hamamatsu R9779 PMT. Detectors
are operated at \textminus{}1300 V and the external CFD delay is set to 1.6 ns.
The zero time references is given by the position of 1173-keV full energy peak.}
\end{figure}

Obviously the Z value affects the time walk, which needs to be corrected for when the
detectors are used in a real experiment in a wider energy range. Nevertheless the Compton time walk provides a good approximation of the FEP time walk when using small crystals like those under study in this paper \cite{Mach1991197}. 
In order to define suitable operational values, it is therefore useful to examine the time position as a function of energy for the different settings . Figure \ref{fig:walk curve} shows the centroid position
of the time distribution as a function of energy for Compton events arising from the 1173-keV 
$\gamma$-ray from $^{60}$Co, for Crystals $A$ (standard) and $B$ (enhanced doping of 8\%). 
As depicted in the figure positive values of Z yield a smooth behaving time walk, which can 
be accounted for in the measurements. 
At the best Z value for time resolution, the Compton walk is of the order of 300 ps over 
1 MeV for Crystal $B$.
Although smaller walk can be obtained at a setting of Z=1.0 mV, this is at the expense of a steeper
curve at low energies. 
Crystal $A$ has a similar behaviour, but in this case it is possible to achieve walk of the order of 150 ps/MeV 
for settings Z=1.0 mV or Z=0.5 mV.

\section{Summary and conclusions}

We have investigated the time response of three scintillator detectors equipped with
cylindrical LaBr$_{3}$(Ce) crystals, 1-in in height and 1-in. in diameter, produced with 
different nominal Ce doping.
The aim of the study was to obtain the best time resolution through the optimization
of the electronics parameters, while keeping the detectors good energy resolution and linearity.
The crystals were tested with the Hamamatsu R9779 photomultiplier tube,
which was selected due to its good timing properties, against a fast BaF$_{2}$ reference detector, at
1173-keV/1332-keV ($^{60}$Co) and 511-keV ($^{22}$Na) energies. Time signals were processed with analogue electronics
by means of a 935 CFD and a 567 TAC by ORTEC. The CFD parameters were scanned
and fine-tuned in order to achieve the narrowest time distribution.

We report the time resolutions at $^{60}$Co energies, given as 
the FWHM for the individual \labr\ detector, of 106\textpm{}2, 98\textpm{}2 and 107\textpm{}2 ps, 
for Crystals $A$ (standard 5\% Ce doping), $B$ (8\% Ce) and $C$ (test crystal 10\% Ce),
respectively.
The differences show the effect of the amount of Ce and the crystal structure.
As expected the time resolution of 
of Crystal $B$ (8\% Ce) is better than Crystal $A$ (5\% Ce) \cite{Glodo2005}, but, 
in contrast to former studies,  the time resolution of Crystal $C$ (10\% Ce) 
is worse. This could be understood as inhomogeneities in the doping distribution
at the time when it was manufactured as a test crystal. 

We also conclude that the leading edge operation mode
of the ORTEC 935 may provide similar or better values for the time resolution
than the CFD operation method at a given energy. This is the case of Crystal $A$
after the optimization of the triggering fraction by the modification of the threshold
control value of the CFD module. 
For the operation in the CFD regime, the time walk has been assessed by taking 
advantage of the good ORTEC 935 qualities and the possibility of precise zero-crossing
adjustment. We have shown that a very smooth time walk is obtained for energies
lower than 1 MeV. 

Finally, for Crystal $B$, with enhanced Ce doping of 8\%, we have found that a bias voltage of 
\textminus{}1300 V on the PMT, together with the optimal CFD parameters of Z= 2.0 mV and external 
delay of 1.6 ns, yield the best time resolutions at $^{60}$Co and $^{22}$Na energies. 
The time resolution measured at $^{60}$Co energies has been pushed below 100 ps 
to 98$\pm$2 ps, which should be compared to the best resolution available to date 
with a similar cylindrical 1-in. LaBr$_{3}$(Ce) crystal of 107\textpm{}4
ps, reported by Moszy\'{n}ski \textit{et al.}, using the Photonis XP20D0 PMT \cite{Mos2006}.
This improved time resolution represents and advancement for the application of the
fast-timing ATD method.

In conclusion we have shown that, apart from crystal and photomultiplier
selection, the optimization of electronic parameters leads to improved 
time resolution for standard 1-in. LaBr$_{3}$(Ce) detectors. 
This work will be extended further by studying larger crystals and 
different geometries, which are of relevance for the construction of the FAst TIMing Array (FATIMA)
for studies of nuclei far off the line of stability. 

\section*{Acknowledgements}

\label{ackn} 
This work was partially supported by the Spanish MINECO through projects FPA2010-17142, FPA2013-41267-P and Consolider-CPAN CSD-2007-00042. 
Funding by the ERA-NET NuPNET via Spanish project PRI-PIMNUP-2011-1338 (FATIMA-NuPNET) and Bulgarian project DNS7RP01-4  (FATIMA-NuPNET) is 
also recognized. V.V. and H.M. acknowledge the financial aid by the Consolider-CPAN CSD-2007-00042 project. The electronics for the test bench and the reference detectors were provided by the Fast Timing Pool of Electronics and MASTICON.

\section*{Bibliography}
\bibliographystyle{elsarticle-num}
\bibliography{labr3_1inch}

\end{document}